\documentclass[aps,prl,reprint,superscriptaddress]{revtex4-2}

\usepackage{amsmath,amssymb,graphicx,bm}
\usepackage[colorlinks=true,linkcolor=blue,citecolor=blue,urlcolor=blue]{hyperref}
\usepackage{graphicx}
\usepackage{xcolor}

\begin{document}

\title{Termination-Controlled Fractionalization and Hybridization at Topological Interfaces in Organic Spin Chains}

\author{Khalid N. Anindya}
\email{Corresponding author: khalid.anindya@mcgill.ca}
\author{Hong Guo}
\affiliation{Department of Physics, McGill University, Montréal, QC H3A 2T8, Canada}

\date{\today}

\begin{abstract}
A single organic spin platform hosts both dimerized \(S=\tfrac{1}{2}\) and effective Haldane \(S=1\) sectors, linked by bond-texture inversion. At the junction, the fractional mode is controlled by termination parity: quenched by local fusion at one termination and released as an uncompensated spin-\(\tfrac{1}{2}\)-like degree of freedom at the parity-shifted one. Two such internal boundary modes of a finite embedded Haldane domain hybridize with an exponentially decaying splitting, establishing termination parity as a design principle for engineering and coupling fractional boundary modes.
\end{abstract}

\maketitle

\section{Introduction}

Symmetry-protected topological (SPT) phases in one dimension are distinguished by a gapped bulk and emergent boundary degrees of freedom that cannot be removed without either closing the gap or breaking the protecting symmetry.\cite{Haldane1983a,Haldane1983b,AKLT1987,AKLT1988,Pollmann2010,Pollmann2012} Among their canonical realizations, the spin-$1$ Haldane chain supports fractional spin-$\tfrac12$ end states and nonlocal string order, while bond-alternating spin-$\tfrac12$ chains furnish an interacting analogue of the Su--Schrieffer--Heeger (SSH) problem in which the topological sector is selected by the termination pattern.\cite{SSH1979,Heeger1988,Hida1992,Wang2013,denNijs1989} In both cases, the physically relevant objects are not single-particle band edge states but many-body boundary excitations rooted in quantum entanglement.\cite{Pollmann2010,Pollmann2012}

A junction between two distinct one-dimensional topological spin sectors is, in principle, expected to involve a fractional spin-$\tfrac12$ boundary degree of freedom.\cite{Chepiga2016,Chepiga2016b,Kenzelmann2003} However, for a realistic heterojunction this abstract existence statement is not yet the central physical question. What matters experimentally is whether the interfacial fractional mode remains \emph{observable} as a localized low-energy degree of freedom, or whether it is instead \emph{locally quenched} by fusion with a compensating boundary contribution from the adjoining segment. At a junction between a Haldane $S=1$ chain and a dimerized $S=\tfrac12$ chain, this visibility problem is controlled by boundary bookkeeping: specifically, by the termination parity of the dimerized side at the interface. This distinction goes beyond bulk or solitonic treatments of the Haldane--dimerized boundary.\cite{Chepiga2016,Chepiga2016b}  Such analyses establish that fractional spin-$\tfrac12$ objects can arise at phase boundaries, but they do not address the sharper heterojunction question relevant here: when two inequivalent SPT chain architectures are joined spatially, under what conditions is the interfacial fractional mode released as an active localized boundary excitation, and under what conditions is it fused away locally? In the non-interacting SSH analogue, it is precisely the boundary termination that determines whether an edge zero mode is present.\cite{SSH1979,Heeger1988} The many-body analogue of this \emph{termination-controlled visibility} at a Haldane/dimerized heterojunction, to our knowledge, has not been demonstrated in a chemically grounded platform.

Recent advances in bottom-up nanographene synthesis and scanning-probe spectroscopy have made such interacting one-dimensional spin systems experimentally accessible with atomic precision.\cite{Mishra2021,Zhao2024,Zhao2025,Henriques2025dispersion} Fractional edge excitations have been observed directly in nanographene spin chains,\cite{Mishra2021} and tunable exchange-alternating spin-$\tfrac12$ Heisenberg chains have now been realized on surfaces.\cite{Zhao2024} At the same time, open-shell triangulene derivatives and related aza-triangulene motifs provide a chemically versatile route to robust $\pi$-radicals and designer exchange pathways in organic platforms.\cite{Wang2022AzaTriangulene,Pawlak2025KagomeRadicals} These developments raise a natural next question: beyond identifying isolated endpoint phases, can one engineer \emph{internal topological interfaces} within a single chemically realistic platform and control whether their fractional boundary modes are quenched or released?

In our recent ACS Nano work, we showed that a trioxoazatriangulene (TANGO)-based one-dimensional Mott chain supports precisely such a dual setting: selective radicalization of the same molecular backbone yields either a bond-centered dimerized spin-$\tfrac12$ chain or a Hund-coupled effective Haldane spin-$1$ chain, both deep in the Mott regime and both exhibiting clear many-body SPT diagnostics.\cite{Anindya2026ACSNano} That work established the endpoint phases using quantized many-body Zak phases, entanglement fingerprints, edge multiplets, and real-space boundary magnetization, demonstrating that a single organic scaffold can host two inequivalent one-dimensional topological spin architectures.\cite{Anindya2026ACSNano} What remained open, however, was the interfacial problem: once these two sectors are embedded within a common parent-family setting, what many-body boundary physics appears at the junction between them? 

Here we resolve this question in a concrete many-body setting by introducing a TANGO-based organic parent spin-chain family that supports two inequivalent effective topological architectures: a dimerized spin-$\tfrac12$ chain and a Hund-coupled effective Haldane spin-$1$ chain.\cite{Anindya2026ACSNano} 
We show that a single junction between the two sectors exhibits \emph{termination-controlled fractionalization}: in one interface geometry, neighboring spin-$\tfrac12$ boundary contributions fuse and quench the visible interfacial response, whereas in a parity-shifted geometry an uncompensated interfacial spin-$\tfrac12$-like mode is unmasked. This termination-controlled fractionalization is distinct from
the bulk soliton picture\cite{Chepiga2016b} and
from prior interface constructions in non-interacting
systems:\cite{SSH1979,Heeger1988} it is a property of the full
many-body boundary-mode manifold, quantified by the
integrated local spin saturating to $\tfrac{1}{2}$
at the active interface and remaining zero at the quenched one. Finally, by embedding a finite effective Haldane domain inside non-SPT-terminated dimerized regions, we isolate two active internal interfaces and show that their low-energy splitting decays approximately exponentially with separation, identifying the splitting as the hybridization of two localized spin-$\tfrac12$ boundary modes of the embedded Haldane segment. In a bottom-up realized chain architecture, local \(dI/dV\) spectroscopy should be able to distinguish the active junction from the locally quenched one through the presence or absence of a spatially localized zero-bias feature,\cite{Mishra2021,Zhao2024,Henriques2025dispersion} and to track the controlled hybridization splitting of embedded Haldane domains, providing a direct experimental signature of termination-controlled fractionalization and interface-mode coupling. These results elevate the platform from a host of isolated SPT phases to a controllable setting for interfacial fractionalization and boundary-mode coupling in organic quantum spin chains.

\section{Model and Numerical Methods}

We model the system as a nearest-neighbor spin-$\tfrac12$ Heisenberg chain,
\begin{equation}
H=\sum_{i=1}^{L-1} J_i\,\mathbf{S}_i\!\cdot\!\mathbf{S}_{i+1},
\label{eq:H_parent}
\end{equation}
with alternating bond families $J_1$ and $J_2$. The two endpoint coupling sets correspond to the dimerized spin-$\tfrac12$ and Hund-coupled effective Haldane limits previously extracted for the same TANGO platform from DFT$+U$ exchange mapping.\cite{Anindya2026ACSNano} To establish that these endpoints belong to a common parent family, we introduce an effective interpolation $J_{1,2}(\lambda)$ between them; this $\lambda$ is used only to diagnose the bulk bond-texture inversion, whereas all single-interface and two-interface calculations are performed directly at the endpoint couplings. Ground-state and low-energy properties were computed by finite-system DMRG with explicit $U(1)$ conservation of $S_{\mathrm{tot}}^z$, and benchmarked against iDMRG in the bulk limit.
Full details of the Hamiltonian construction, sector
assignments, interface geometries, and fitting procedures
are given in Secs.~S1--S8 of the Supplemental Material (SM).\cite{SM}

\begin{figure*}[!ht]
\includegraphics[width=\textwidth]{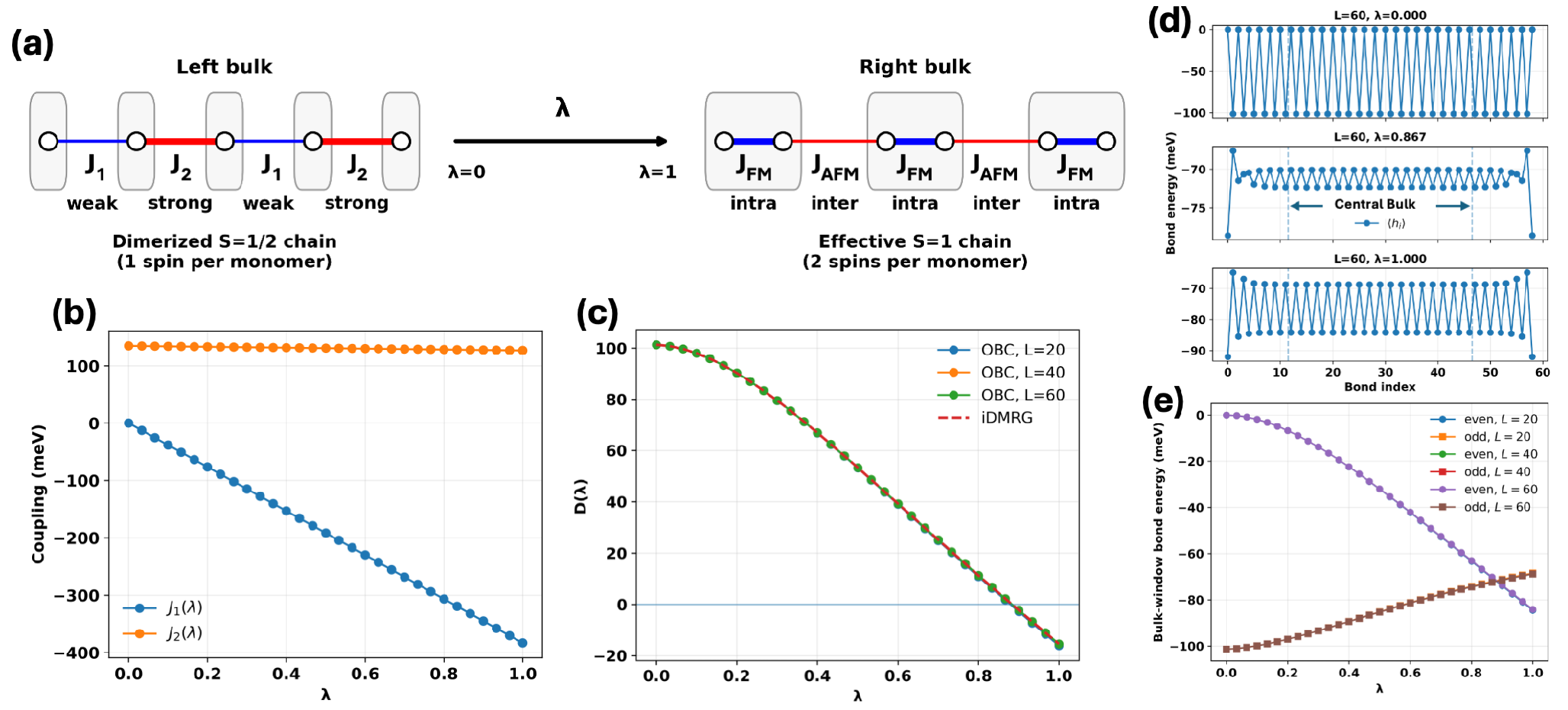}
\caption{\textbf{Effective parent-family interpolation between dimerized \(S=\tfrac12\) and effective Haldane \(S=1\) chain architectures.}
(a) Schematic of the two endpoint
descriptions and their connection through the interpolation
parameter $\lambda$: a dimerized $S=\tfrac{1}{2}$ chain
with alternating couplings $(J_1, J_2)$, and an effective
Haldane $S=1$ chain with intra-monomer Hund coupling
$(J_{\rm FM})$ and inter-monomer antiferromagnetic exchange
$(J_{\rm AFM})$. (b) Interpolating couplings \(J_1(\lambda)\) and \(J_2(\lambda)\). (c) Bulk-window dimerization \(D(\lambda)\), showing a robust sign change along the interpolation and good agreement between OBC and iDMRG. (d) Representative local bond-energy profiles \(\langle h_i\rangle\) for three values of \(\lambda\). (e) Bulk-window even- and odd-bond energies, showing that the sign change of \(D(\lambda)\) arises from a genuine reordering of the two bond sectors.}
\label{fig:fig1}
\end{figure*}

\section{Results and Discussion}

\subsection{Bulk parent-family evolution and bond-texture inversion}

Building on the endpoint dimerized and Hund-coupled limits established previously,\cite{Anindya2026ACSNano} we first examine the bulk evolution of the parent-family chain before introducing spatial interfaces between the two sectors. 
Since $\lambda$ serves only to establish the bulk
bond-texture inversion (Sec.~S1~\cite{SM}), we focus
here on its physical consequences before turning to
the interface calculations.

Fig.~\ref{fig:fig1}(a) summarizes the physical content of the two endpoint descriptions. At \(\lambda=0\), the low-energy degrees of freedom form a dimerized spin-$\tfrac12$ chain with alternating weak and strong neighboring-monomer couplings \(J_1\) and \(J_2\). At \(\lambda=1\), the same parent-family notation evolves into a Hund-coupled regime in which \(J_1<0\) locks two spin-$\tfrac12$ moments within a monomer into an effective triplet, while \(J_2>0\) couples neighboring effective spin-$1$ units antiferromagnetically.

As shown in Fig.~\ref{fig:fig1}(b), increasing \(\lambda\) drives \(J_1\) from a weak antiferromagnetic bond to a strong Hund-coupled ferromagnetic bond, while \(J_2\) remains antiferromagnetic and varies comparatively little. This already suggests that the dominant bond hierarchy must reorganize across the interpolation, but the physically relevant quantity is the resulting many-body bond texture rather than the bare couplings themselves.

To probe this, we define central-window even- and odd-bond averages, \(\overline{E_{\rm even}}\) and \(\overline{E_{\rm odd}}\), over the region marked in Fig.~\ref{fig:fig1}(d) (see Sec.~S3\cite{SM}).
The resulting bulk-window dimerization ($D(\lambda)=\overline{E_{\rm even}}-\overline{E_{\rm odd}}$) is shown in Fig.~\ref{fig:fig1}(c). Several open-boundary chain lengths collapse onto essentially the same curve, indicating that the bulk-window construction efficiently removes edge contributions and that the bond-texture evolution is already well converged at the system sizes studied here. The iDMRG benchmark tracks the same behavior closely, confirming that the sign change of \(D(\lambda)\) is a genuine bulk feature rather than a boundary artifact. The important point is not only that \(D(\lambda)\) changes magnitude, but that it changes \emph{sign}, signaling an inversion of the dominant bond sector along the parent-family evolution.

The real-space origin of this inversion is clarified by the representative bond-energy profiles in Fig.~\ref{fig:fig1}(d). Deep in the dimerized regime, the chain exhibits a pronounced alternating weak/strong pattern characteristic of the spin-$\tfrac12$ endpoint. As \(\lambda\) is increased toward the inversion region, the contrast between the two bond sectors is strongly reduced in the bulk, indicating a reorganization of the underlying bond hierarchy. Upon entering the Hund-coupled side, a different alternating pattern emerges, now associated with the effective Haldane architecture. Thus, the parent-family interpolation does not simply renormalize an existing dimer pattern; it drives a qualitative rearrangement of the bulk bond texture.

This interpretation is sharpened further in Fig.~\ref{fig:fig1}(e), where the even- and odd-bond energies are plotted separately. Their crossing shows that the sign change of \(D(\lambda)\) is not a subtraction artifact, but reflects a genuine reordering of the two bond-energy branches. The bulk evolution is therefore controlled by an inversion of bond-sector dominance between the two endpoint architectures. Figure \ref{fig:fig1} thus establishes the bulk parent-family setting from which the interface physics emerges.

\begin{figure*}[!ht]
\includegraphics[width=\textwidth]{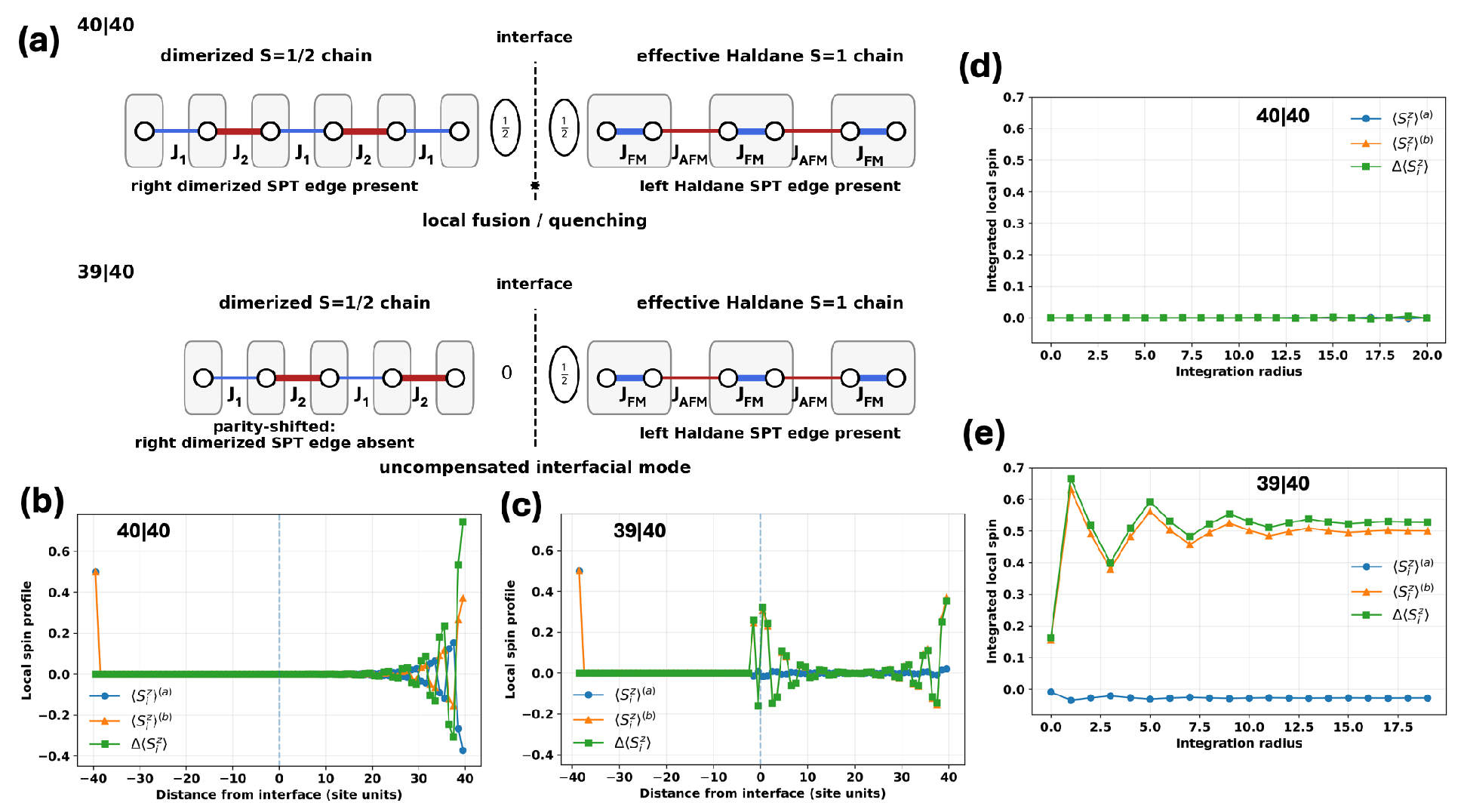}
\caption{\textbf{Termination-controlled fractionalization at a single interface.}
(a) Single-interface constructions for the \(40|40\) and parity-shifted \(39|40\) chains. In \(40|40\), the dimerized and Haldane sides both contribute spin-\(\tfrac12\) SPT boundary moments at the junction, allowing local fusion and quenching. In \(39|40\), the dimerized-side contribution is removed, leaving an uncompensated interfacial spin-\(\tfrac12\)-like mode on the Haldane side. (b,c) Site-resolved spin profiles for two selected low-energy states, \(\langle S_i^z\rangle^{(a)}\) and \(\langle S_i^z\rangle^{(b)}\), together with \(\Delta\langle S_i^z\rangle=\langle S_i^z\rangle^{(b)}-\langle S_i^z\rangle^{(a)}\), plotted versus distance from the interface for \(40|40\) and \(39|40\), respectively. (d,e) Integrated local spin in a window of radius \(r\) centered on the interface for the same two geometries. The \(40|40\) interface remains essentially spin-quiet, whereas the \(39|40\) interface builds up and saturates near \(1/2\), demonstrating termination-controlled interfacial fractionalization.}
\label{fig:fig2}
\end{figure*}

\subsection{Termination-controlled fractionalization at a single interface}

We now turn to the question of what occurs when the two endpoint sectors are joined spatially within a single chain. To isolate this physics, we construct sharp single-interface geometries directly at the bond level, with a left dimerized spin-$\tfrac12$ segment joined to a right Hund-coupled effective Haldane segment. Here and below, the notation \(A|B\) denotes a geometry consisting of a left dimerized segment of \(A\) spin-$\tfrac12$ sites joined to a right effective Haldane segment of \(B\) spin-$\tfrac12$ sites; changing \(A\) by one site changes the termination parity on the dimerized side. The key point is that the local interface response is controlled not only by the bulk identities of the two adjoining regions, but also by how their boundary bookkeeping is matched at the junction.

This is illustrated by the two constructions in Fig.~\ref{fig:fig2}(a). In the \(40|40\) chain, the right edge of the left dimerized segment is \(J_1\)-terminated and therefore contributes a spin-$\tfrac12$ SPT boundary moment, while the left edge of the effective Haldane segment contributes its characteristic spin-$\tfrac12$ boundary mode. These two fractional contributions meet at the junction and can fuse locally, suppressing the visible interfacial response. In the parity-shifted \(39|40\) chain, by contrast, the right edge of the dimerized segment becomes \(J_2\)-terminated and no longer contributes an interface-side SPT spin. The Haldane-side boundary mode is then left uncompensated, allowing the interface itself to carry a localized spin-$\tfrac12$-like response.

The site-resolved spin textures in Fig.~\ref{fig:fig2}(b,c) make this distinction explicit. For the \(40|40\) geometry, the selected low-energy profiles \(\langle S_i^z\rangle^{(a)}\), \(\langle S_i^z\rangle^{(b)}\), and their difference \(\Delta\langle S_i^z\rangle\) show that the dominant response is concentrated away from the junction and is instead governed by the outer edge degrees of freedom. In particular, the right open end of the Haldane segment retains the characteristic localized edge texture, while the interface itself remains comparatively spin-quiet, consistent with local fusion of the two boundary contributions. In the \(39|40\) geometry, however, the corresponding \(\Delta\langle S_i^z\rangle\) profile is centered at the interface, showing that the parity shift releases an uncompensated interfacial fractional mode rather than merely reorganizing ordinary outer-edge weight. A more detailed interpretation of the sector-resolved spin textures, together with additional bond-energy and entanglement-entropy diagnostics for the \(40|40\) and \(39|40\) geometries, is provided in Secs.~S9--S10 of the SM (see also Fig.~S1). \cite{SM} 

\begin{figure*}[!ht]
\includegraphics[width=\textwidth]{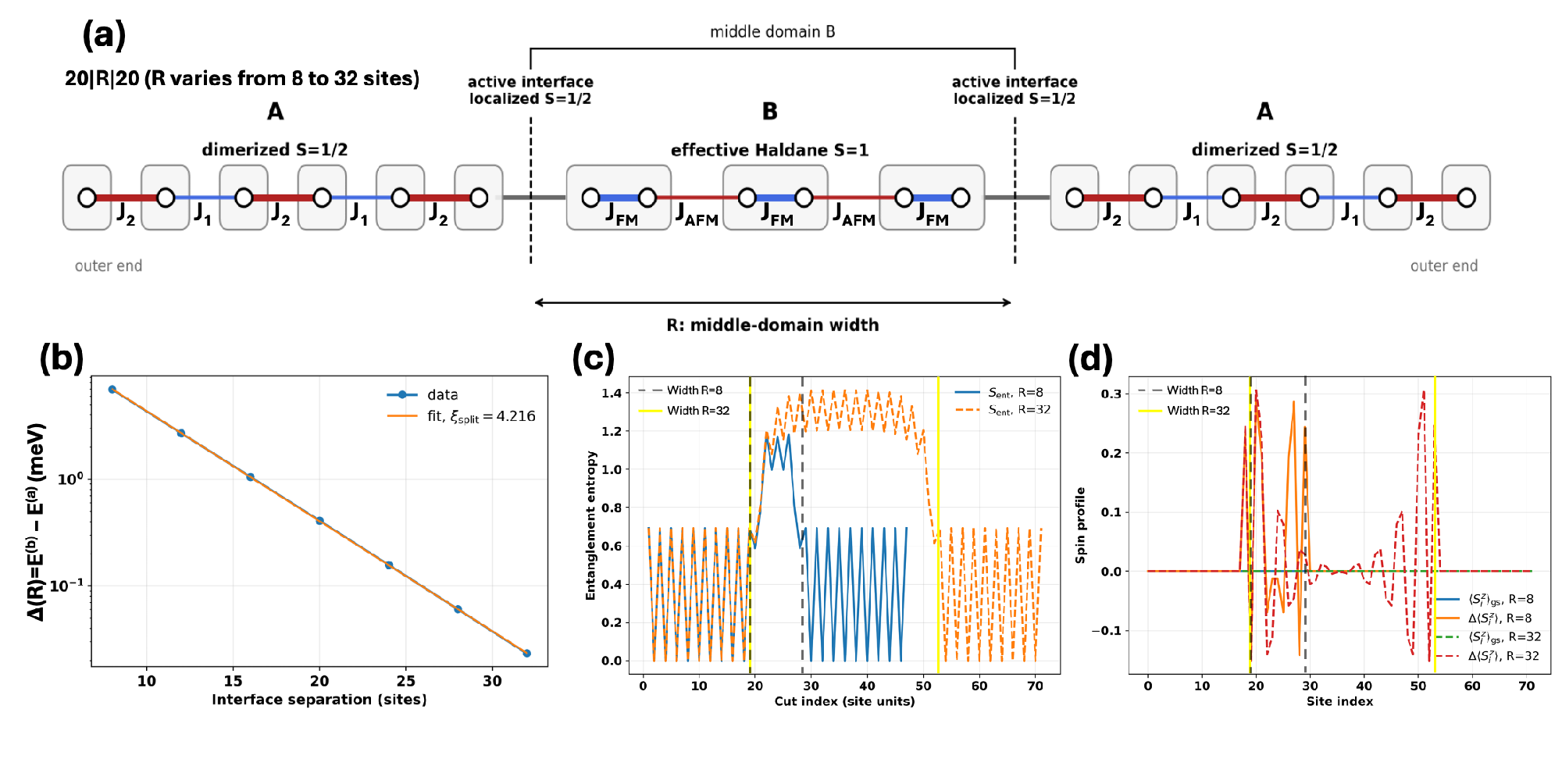}
\caption{\textbf{Exponential hybridization of two internal interface modes.}
(a) Two-interface \(A|B|A\) construction used to isolate internal boundary modes. The outer dimerized ends are chosen to be non-SPT (\(J_2\)-terminated), so that the low-energy physics is dominated by the two active internal interfaces between the dimerized \(A\) regions and the embedded effective Haldane \(B\) region. The middle-domain width \(R\) sets the interface separation. (b) Low-energy splitting \(\Delta_{ST}=|E^{(b)}-E^{(a)}|\) versus interface separation, showing an approximately exponential decay. (c) Entanglement-entropy profiles for representative small and large separations. (d) Representative real-space spin profiles for small and large \(R\), showing strong overlap at short separation and two resolved localized packets at larger separation. The $R=8$ and $32$ spin-\(\tfrac12\) sites window are marked by dashed grey and solid yellow lines. These data identify the splitting as the exponential hybridization of the two spin-\(\tfrac12\) SPT boundary modes of the embedded Haldane domain.}
\label{fig:fig3}
\end{figure*}

This qualitative contrast is quantified directly by the integrated local spin in Fig.~\ref{fig:fig2}(d,e). For the \(40|40\) chain, the spin accumulated in a window of radius \(r\) centered on the interface remains essentially zero, confirming that the junction is locally quenched despite the presence of fractional boundary degrees of freedom elsewhere in the chain. By contrast, for the \(39|40\) chain the integrated interface spin grows rapidly with \(r\) and saturates near \(1/2\), directly demonstrating an uncompensated localized fractional moment. The contrast between the two cases therefore identifies termination-controlled interfacial fractionalization. 

Figure \ref{fig:fig2} thus establishes the central interface principle of the present work: a junction between the dimerized spin-$\tfrac12$ and effective Haldane spin-$1$ sectors does not host a universal visible boundary mode. Instead, the interface response depends on whether the adjoining fractional boundary contributions fuse and quench locally, or whether one side is termination-suppressed so that an uncompensated interfacial spin-$\tfrac12$-like mode is released. This provides the conceptual bridge from the bulk parent-family evolution to the two-interface hybridization physics discussed next.

\subsection{Exponential hybridization of two internal interface modes}

Having established that a single junction can either quench or release an interfacial fractional mode depending on its termination bookkeeping, we now ask how two such active interfaces interact when brought into the same chain. To isolate this physics, we construct explicit \(A|B|A\) geometries in which a finite effective Haldane domain \(B\) is embedded between two dimerized outer regions \(A\). The outer dimerized ends are chosen to be non-SPT (\(J_2\)-terminated), so that the low-energy response is governed predominantly by the two internal interfaces rather than by physical chain ends.

The construction is shown in Fig.~\ref{fig:fig3}(a). Each \(A|B\) boundary hosts a localized spin-\(\tfrac12\)-like mode associated with the boundary of the embedded Haldane segment. If these two internal modes are well separated, they should behave as nearly independent localized objects; if they are brought closer together, their overlap should generate a finite hybridization splitting. The natural control parameter is therefore the interface separation \(R\), defined by the width of the embedded Haldane domain. This mechanism is closely analogous to the familiar finite-size singlet--triplet splitting of an open Haldane chain, which likewise reflects overlap between two localized spin-\(\tfrac12\) boundary modes.\cite{AKLT1987,Kenzelmann2003} The crucial difference here is that the relevant modes reside at engineered \emph{internal} topological interfaces of an embedded Haldane domain, while the physical outer ends are rendered non-SPT.

Fig.~\ref{fig:fig3}(b) shows the resulting low-energy splitting
\begin{equation}
\Delta(R) \equiv E^{(b)}-E^{(a)}
\end{equation}
as a function of \(R\). On a linear scale, the splitting is largest at short separation and decreases rapidly as the interfaces are moved apart. More importantly, on a semilogarithmic scale the data are approximately linear over the accessible range, consistent with
\begin{equation}
\Delta(R)\sim e^{-R/\xi_{\rm split}}.
\end{equation}
This is the characteristic signature of two localized boundary modes whose overlap decays exponentially with distance. The extracted \(\xi_{\rm split}\) therefore provides an energy-space measure of the localization length of the internal interface states.

The corresponding many-body correlation structure is illustrated in Fig.~\ref{fig:fig3}(c), which shows representative bipartite entanglement-entropy profiles for small and large \(R\). At short separation, the entanglement reconstruction associated with the two interfaces strongly overlaps in the middle of the chain. As \(R\) increases, it evolves into two more weakly overlapping interface-centered structures, consistent with progressive spatial separation of the two localized modes.

This interpretation is confirmed directly by the real-space spin textures in Fig.~\ref{fig:fig3}(d). For small \(R\), the spin response forms a single broad central structure, indicating substantial overlap between the two internal interface modes. For larger \(R\), the response resolves into two distinct localized packets centered at the two interfaces. The lower state remains only weakly polarized, while the higher state carries the visible interface-centered spin weight. The evolution from a merged central texture to two separated interface packets mirrors the exponential collapse of \(\Delta(R)\), showing that the low-energy splitting originates from hybridization of internal interfacial modes rather than from residual outer-edge contributions. Consistently, the polarized weight remains centered at the two internal interfaces, while the physical chain ends stay essentially spin-silent throughout, confirming that the splitting is controlled by the
internal topology, not by open-boundary effects. A more detailed discussion of the sector-resolved spin textures, together with additional linear-scale splitting data and representative bond-energy profiles for the two-interface geometries, is provided in Secs.~S11--S12 of the SM (see also Fig.~S2). \cite{SM}

Figure \ref{fig:fig3} therefore completes the interface story initiated above. Once termination-controlled fractionalization is used to isolate active internal junctions, the resulting localized spin-\(\tfrac12\)-like modes behave as bona fide boundary degrees of freedom of the embedded Haldane domain: when the interfaces are close, they hybridize strongly, whereas for larger separations their coupling collapses approximately exponentially.

\section{Conclusion}

We have shown that a single organic spin-chain platform provides
a controlled many-body setting for topological interface physics
between a dimerized $S=\tfrac{1}{2}$ and an effective Haldane
$S=1$ sector.
The two endpoints, both realizable on the same molecular scaffold, are connected through a
bulk bond-texture inversion within one parent family, establishing
that their junction is a true topological interface rather than an
accidental boundary. 
At that junction, the interfacial response is not universal: depending on the boundary bookkeeping, the fractional mode is either fused away locally or released as an uncompensated spin-$\tfrac12$-like degree of freedom. For a finite embedded Haldane domain, the resulting two active internal boundary modes hybridize with an approximately exponential dependence on separation. Our results therefore identify termination parity as a practical design principle for creating, quenching, releasing, and coupling internal fractional boundary modes within a single organic spin-chain platform.

\section{Acknowledgements}

We gratefully acknowledge financial support from Natural Sciences and Engineering Research Council of Canada (NSERC) and Fonds de recherche du Québec – Nature et technologies (FRQNT). The authors thank the Digital Research Alliance of Canada and Calcul Québec for substantial
computational allocations that made this work possible.

\bibliographystyle{apsrev4-2}

\end{document}


\title{Supplemental Material for: \textit{Termination-Controlled Fractionalization and Hybridization at Topological Interfaces in Organic Spin Chains}}

\author{Khalid N. Anindya}
\email{Corresponding author: khalid.anindya@mcgill.ca}
\author{Hong Guo}
\affiliation{Department of Physics, McGill University, Montréal, QC H3A 2T8, Canada}

\date{\today}

\maketitle

\section{S1. Parent-family Hamiltonian and interpolation}

The calculations in the main text are based on an explicit nearest-neighbor spin-$\tfrac12$ Heisenberg chain,
\begin{equation}
H=\sum_{i=1}^{L-1} J_i\,\mathbf{S}_i\!\cdot\!\mathbf{S}_{i+1},
\label{eq:SM_H_parent}
\end{equation}
where even bonds correspond to the intra-monomer channel and odd bonds correspond to the inter-monomer channel. We emphasize that all physical sites remain spin-$\tfrac12$ degrees of freedom throughout; the effective Haldane regime emerges dynamically from the hierarchy \(J_1<0\), \(|J_1|\gg J_2>0\), rather than from a coarse-grained rigid spin-$1$ model. This choice preserves the internal structure of the Hund monomers and follows the microscopic description established in Ref.~\onlinecite{Anindya2026ACSNano}.

The two endpoint coupling sets are taken from the DFT\(+U\) plus exchange-mapping analysis of Ref.~\onlinecite{Anindya2026ACSNano}:
\begin{align}
J_1^{\rm dim}&=0.11~{\rm meV}, &
J_2^{\rm dim}&=135.14~{\rm meV},\\
J_1^{\rm Hund}&=-383.19~{\rm meV}, &
J_2^{\rm Hund}&=127.08~{\rm meV}.
\end{align}
To connect these two regimes within one parent family, we use the phenomenological interpolation
\begin{align}
J_1(\lambda)&=(1-\lambda)J_1^{\rm dim}+\lambda J_1^{\rm Hund},\\
J_2(\lambda)&=(1-\lambda)J_2^{\rm dim}+\lambda J_2^{\rm Hund},
\end{align}
with \(0\le\lambda\le1\). 

\(\lambda\) parametrizes an effective interpolation between two low-energy chain architectures (as shown in Eqs.~2--5).
We emphasize that $\lambda$ is introduced solely as a
theoretical device to establish that the two endpoint descriptions belong to a connected
parent family and that varying between them drives a genuine
bulk bond-texture inversion.
It is not proposed as a physical control parameter:
all interface and hybridization calculations presented
below are performed directly at the two endpoint coupling
sets of Eqs.~2 and 3, which are fixed by the
DFT$+U$ analysis of the TANGO platform.~\cite{Anindya2026ACSNano}
The parent-family scan therefore serves only to confirm
that the two sectors share a common microscopic ancestry
and that the bulk separating them undergoes a well-defined
topological reorganization.

For uniform chains, the bond list is generated as
\begin{equation}
J_{2n}=J_1(\lambda),\qquad
J_{2n+1}=J_2(\lambda),
\end{equation}
for a chain of \(L\) spin-$\tfrac12$ sites. For inhomogeneous chains, a cell-wise \(\lambda\)-profile is first defined and then converted into physical bonds. In the simplest cell-profile construction, intra-cell bonds use the local cell value, while inter-cell bonds use the average of neighboring cell values. This avoids introducing an artificial discontinuity on the link between adjacent cells and keeps the interface parity well defined.

\section{S2. DMRG protocol and sector selection}

All finite-chain calculations were carried out with two-site DMRG using TeNPy.\cite{White1992,Schollwock2011,Hauschild2018} We conserved the \(U(1)\) quantum number \(S_{\rm tot}^z\) and initialized the matrix-product state in the desired symmetry sector from a product state with fixed
\begin{equation}
2S_{\rm tot}^z = N_\uparrow-N_\downarrow.
\end{equation}
The DMRG runs employed a two-site update with a mixer, a maximum bond dimension \(\chi_{\max}\), and a truncation threshold \( {\rm svd\_min}=10^{-10}\). In the production calculations used for the figures, the number of sweeps and \(\chi_{\max}\) were increased until the observables shown in the main text were stable within plotting accuracy.

\section{S3. Bulk observables for the parent-family scan}

The bulk evolution in Fig.~1 of main manuscript is characterized from the local bond energies
\begin{equation}
\langle h_i\rangle = J_i \langle \mathbf{S}_i\!\cdot\!\mathbf{S}_{i+1}\rangle,
\end{equation}
evaluated directly from the DMRG wave function. To suppress boundary effects, all bulk averages are taken over a central window \([i_{\rm min},i_{\rm max}]\) that excludes a fixed fraction of sites near the ends. Inside this window, we define the even- and odd-bond averages
\begin{equation}
\overline{E_{\rm even}}=\frac{1}{N_{\rm even}}\sum_{i\in{\rm even}} \langle h_i\rangle,\qquad
\overline{E_{\rm odd}}=\frac{1}{N_{\rm odd}}\sum_{i\in{\rm odd}} \langle h_i\rangle,
\end{equation}
where the sums run only over bonds inside the central bulk window and \(N_{\rm even}\) and \(N_{\rm odd}\) are the corresponding numbers of even and odd bonds. 
The associated bulk-window dimerization is
\begin{equation}
D(\lambda)=\overline{E_{\rm even}}-\overline{E_{\rm odd}}.
\end{equation}
The sign change of \(D(\lambda)\) is therefore a direct measure of bond-texture inversion in the bulk.

\section{S4. Single-interface constructions}

For the single-interface geometries of Fig.~2 in main manuscript, we construct the Hamiltonian directly at the bond level. A left segment \(A\) occupies sites
\begin{equation}
0,\ldots,L_{\rm left}-1,
\end{equation}
and a right segment \(B\) occupies sites
\begin{equation}
L_{\rm left},\ldots,L-1.
\end{equation}
The interface bond lies between sites \(L_{\rm left}-1\) and \(L_{\rm left}\), and can be assigned independently as either \(J_1\) or \(J_2\), although for the chemistry considered here the natural choice is the inter-monomer bond \(J_2\).

A key feature of this explicit construction is that the internal parity of each segment can be chosen independently. In particular, the left and right segments can be started internally with either \(J_1\) or \(J_2\), allowing clean control over whether a given dimerized edge is \(J_1\)-terminated (SPT-active) or \(J_2\)-terminated (non-SPT). This makes it possible to implement the \(40|40\) and \(39|40\) geometries directly and to isolate the effect of changing the interface-side termination while keeping the bulk couplings fixed.

For a chosen interface position \(x_{\rm I}\), we compute
\begin{equation}
\langle S_i^z\rangle^{(a)},\qquad
\langle S_i^z\rangle^{(b)},\qquad
\Delta \langle S_i^z\rangle=
\langle S_i^z\rangle^{(b)}-\langle S_i^z\rangle^{(a)},
\end{equation}
as well as the integrated local spin
\begin{equation}
\Sigma(r)=\sum_{|i-x_{\rm I}|\le r+1/2}\langle S_i^z\rangle,
\end{equation}
where the half-integer convention ensures that a bond-centered interface includes the two adjacent sites already at \(r=0\). In the main text, the build-up of \(\Sigma(r)\) provides the quantitative distinction between the quenched \(40|40\) interface and the uncompensated \(39|40\) interface.

\section{S5. Two-interface \texorpdfstring{$A|B|A$}{A|B|A} geometries}

For Fig.~3 in main manuscript, we construct an explicit site-resolved \(A|B|A\) geometry, in which a middle Haldane-like region \(B\) is embedded between two outer dimerized regions \(A\). If the left outer segment has length \(L_{\rm A}^{\rm (L)}\), the middle segment has length \(L_{\rm B}\), and the right outer segment has length \(L_{\rm A}^{\rm (R)}\), then the two interface bonds lie at
\begin{equation}
x_{\rm I}^{\rm (L)}=L_{\rm A}^{\rm (L)}-\tfrac12,\qquad
x_{\rm I}^{\rm (R)}=L_{\rm A}^{\rm (L)}+L_{\rm B}-\tfrac12.
\end{equation}
The interface separation is therefore
\begin{equation}
R=x_{\rm I}^{\rm (R)}-x_{\rm I}^{\rm (L)}.
\end{equation}

To isolate the internal interface physics, the physical outer ends of the dimerized segments are chosen to be non-SPT, i.e. \(J_2\)-terminated, so that the dominant low-energy response comes from the two internal interfaces rather than from trivial physical ends. The internal A|B interfaces are kept active, producing two localized spin-$\tfrac12$ boundary modes of the embedded Haldane domain. The low-energy splitting is then defined as
\begin{equation}
\Delta(R)=E^{(b)}-E^{(a)} ,
\end{equation}
and fitted to
\begin{equation}
\Delta(R)=A e^{-R/\xi_{\rm split}}.
\end{equation}
The fitted \(\xi_{\rm split}\) provides an energy-space measure of the interface-mode localization length, which can be compared with the spatial extent of the real-space spin textures.

\section{S6. Entanglement entropy}

For both single- and two-interface calculations, we evaluate the bipartite von Neumann entropy\cite{Vidal2003,Calabrese2004}
\begin{equation}
S(\ell)=-{\rm Tr}\,\rho_\ell \ln \rho_\ell ,
\end{equation}
where \(\rho_\ell\) is the reduced density matrix of the left block after cutting the chain between sites \(\ell\) and \(\ell+1\). This quantity is dimensionless and probes the reorganization of many-body correlations across the cut. In the bulk scan, the entanglement profile is also used to extract an effective central charge through the standard open-chain scaling form,
\begin{equation}
S(\ell)=\frac{c_{\rm eff}}{6}\ln\!\left[\frac{2L}{\pi}\sin\!\left(\frac{\pi \ell}{L}\right)\right]+s_0,
\end{equation}
fitted only over the central cuts inside the chosen bulk window.

\section{S7. Numerical outputs and reproducibility}

For each geometry, the workflow stores the energies, sector labels, bond list, bond-energy profile, spin profile, entanglement profile, and integrated-spin curves in machine-readable tables. In the bulk scan, the output includes the full \(\lambda\)-dependence of \(D(\lambda)\), the separate even- and odd-bond averages, and representative bond profiles. In the interface calculations, the output includes the full bond-resolved and site-resolved data for each geometry, together with the fitted splitting length \(\xi_{\rm split}\) for the two-interface scan. These exports were used directly to generate the publication figures.

\section{S8. Relation to the previous endpoint models}

The present parent-family study does not rederive the endpoint exchange parameters microscopically. Instead, it takes as input the DFT\(+U\)-derived couplings already established in Ref.~\onlinecite{Anindya2026ACSNano}, where the same TANGO platform was shown to realize a dimerized spin-$\tfrac12$ even/odd-Haldane phase and a Hund-coupled effective Haldane spin-$1$ phase. The goal here is different: rather than re-establishing the endpoint SPT classifications, we use those endpoint scales as experimentally grounded anchors and ask what new many-body physics emerges when the two topological sectors are joined through controlled interfaces within a single parent-family description.

\section{S9. Interpretation of the sector-resolved spin textures in the \(40|40\) and \(39|40\) geometries}

The sector-resolved spin textures shown in the main text are most naturally interpreted as representative low-energy configurations of the boundary-mode manifold, rather than as uniquely defined singlet- or triplet-density profiles. In particular, visible local \(\langle S_i^z\rangle\) weight indicates polarization of an emergent fractional mode, but the converse is not true: a boundary mode may exist while carrying little or no visible \(\langle S_i^z\rangle\) in a given sector-selected state.

\paragraph*{\(40|40\) geometry.}
For the \(40|40\) chain, the spin textures indicate that the low-energy manifold is governed primarily by the two \emph{outer} fractional edge moments: the left boundary mode of the dimerized segment and the right boundary mode of the effective Haldane segment. In the representative \(S_{\rm tot}^z=0\) profile, the left outer edge carries positive spin weight while the right outer Haldane edge carries negative spin weight, whereas in the corresponding \(S_{\rm tot}^z=1\) profile the left outer edge remains positively polarized and the right Haldane edge changes sign and becomes positive as well. This is naturally interpreted as an effective two-spin-\(\tfrac12\) manifold: the \(S_{\rm tot}^z=0\) representative corresponds to an antipolarized configuration of the two outer edge moments, while the \(S_{\rm tot}^z=1\) representative corresponds to a parallel configuration. The sign reversal at the right open end therefore does not indicate the appearance of a different localized mode, but rather the same Haldane edge spin-\(\tfrac12\) mode with a different polarization relative to the left outer edge. This behavior supports the view that the interface itself is locally quenched, so that the dominant low-energy response is controlled by the outer-edge degrees of freedom rather than by an active interfacial mode.

\paragraph*{\(39|40\) geometry.}
The \(39|40\) chain is qualitatively different because the parity shift removes the dimerized-side fractional contribution at the junction, leaving the Haldane-side interfacial mode uncompensated. At the same time, the full chain still contains the left outer edge of the dimerized segment and the right open-end edge of the effective Haldane segment. The low-energy manifold is therefore most naturally viewed as involving \emph{three} coupled spin-\(\tfrac12\) boundary degrees of freedom:
\begin{equation}
s_L \;+\; s_I \;+\; s_R,
\end{equation}
where \(s_L\) denotes the left outer dimerized edge moment, \(s_I\) the interfacial Haldane-side boundary mode, and \(s_R\) the right open-end Haldane edge mode.

Within this three-spin picture, it is not necessary that all three fractional modes carry visible local spin weight in the same sector-selected state. In the lower state \(\langle S_i^z\rangle^{(a)}\), the net \(S_{\rm tot}^z=\tfrac12\) can be carried predominantly by the left outer edge \(s_L\), while the two Haldane-side moments \(s_I\) and \(s_R\) remain only weakly polarized or partially compensated. This naturally explains why the right open-end Haldane edge can carry almost no visible weight in \(\langle S_i^z\rangle^{(a)}\), even though in the \(40|40\) geometry the same edge carried a clear opposite-sign response. The difference is not that the right Haldane edge mode disappears, but that the redistribution of polarization within the low-energy manifold is altered once an uncompensated interfacial mode is present.

In the higher state \(\langle S_i^z\rangle^{(b)}\), additional polarization becomes available and is redistributed into the Haldane sector more strongly. As a result, the interfacial spin-\(\tfrac12\)-like mode becomes visible near the junction, and the right Haldane edge can also acquire appreciable weight. The corresponding difference profile,
\begin{equation}
\Delta\langle S_i^z\rangle
=
\langle S_i^z\rangle^{(b)}-\langle S_i^z\rangle^{(a)},
\end{equation}
therefore provides the clearest measure of where the additional spin weight is absorbed. In this sense, the \(39|40\) chain should not be interpreted as showing ``mode absent'' versus ``mode present'' between \(\langle S_i^z\rangle^{(a)}\) and \(\langle S_i^z\rangle^{(b)}\). Rather, both states belong to a low-energy manifold that already contains the uncompensated interfacial degree of freedom, but that mode is only weakly polarized in the lower representative and becomes much more clearly expressed in the higher one.

This interpretation is fully consistent with the integrated local spin shown in the main text: the \(40|40\) geometry remains locally quenched at the interface, whereas the \(39|40\) geometry builds up an interfacial contribution approaching \(1/2\), demonstrating that the parity shift releases an uncompensated localized fractional mode at the junction.

\begin{figure}[!ht]
\centering
\includegraphics[width=\textwidth]{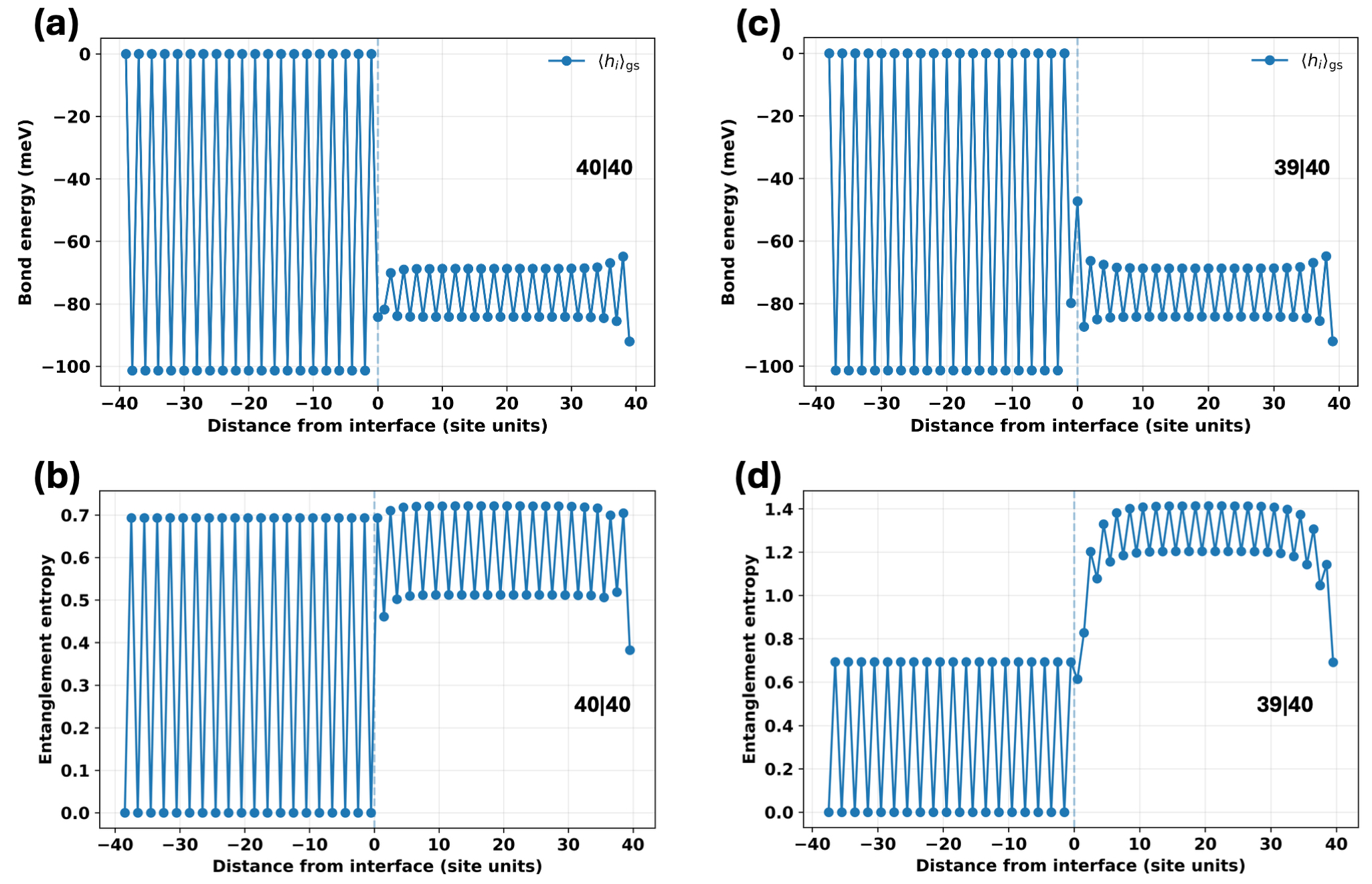}
\caption{\textbf{Additional bond-energy and entanglement diagnostics for the single-interface geometries.}
(a,c) Bond-energy profiles $\langle h_i\rangle_{gs}$ for the \(40|40\) and \(39|40\) chains, respectively, plotted versus distance from the interface. (b,d) Corresponding bipartite entanglement-entropy profiles. The \(40|40\) geometry shows a clear junction between the two bulk bond textures but remains comparatively quiet at the interface, consistent with local quenching. By contrast, the \(39|40\) geometry exhibits a stronger local bond and entanglement reconstruction at the junction, consistent with the release of an uncompensated interfacial fractional mode.}
\label{fig:figs1}
\end{figure}

\section{S10. Additional single-interface diagnostics}

To complement the spin-texture and integrated-spin analysis in the main text, we also examine the bond-energy and entanglement-entropy structure of the two single-interface geometries. Figure \ref{fig:figs1} compares the \(40|40\) and \(39|40\) chains using the local bond energy
\begin{equation}
\langle h_i\rangle = J_i \langle \mathbf{S}_i\!\cdot\!\mathbf{S}_{i+1}\rangle
\end{equation}
and the bipartite entanglement entropy across each bond cut.

For the \(40|40\) geometry, the bond-energy profile in Fig.~\ref{fig:figs1}(a) shows a sharp reconstruction at the junction between the dimerized and effective Haldane sectors, but without producing an anomalously strong localized interface bond. This is consistent with the main-text conclusion that the interface-side fractional contributions are locally fused and that the dominant low-energy response is governed by the outer edge degrees of freedom. The corresponding entanglement-entropy profile in Fig.~\ref{fig:figs1}(b) shows a clear change in the correlation structure across the junction, reflecting the transition from the dimerized side to the effective Haldane side, but without an additional strong interface-localized entanglement enhancement.

The parity-shifted \(39|40\) geometry behaves differently. As shown in Fig.~\ref{fig:figs1}(c), the bond-energy profile develops a more pronounced local reconstruction at the junction, consistent with the release of an uncompensated interfacial mode once the dimerized-side interface contribution is removed. The corresponding entanglement profile in Fig.~\ref{fig:figs1}(d) displays a substantially larger interfacial reorganization than in the \(40|40\) case, again consistent with the emergence of an active localized boundary degree of freedom at the junction. These bond-energy and entanglement diagnostics therefore support the main-text interpretation of Fig.~2: the \(40|40\) interface is locally quenched, whereas the \(39|40\) geometry releases an uncompensated interfacial spin-\(\tfrac12\)-like mode.

\section{S11. Interpretation of the sector-resolved spin textures in the two-interface geometry}

The sector-resolved spin textures of the \(A|B|A\) chain should be interpreted as representative low-energy configurations of the two-interface boundary-mode manifold, rather than as uniquely defined spin densities of isolated local moments. In particular, a localized interface mode may exist while carrying little or no visible \(\langle S_i^z\rangle\) in one selected low-energy state.

For the two-interface geometries considered in the main text, the embedded effective Haldane domain supports two spin-$\tfrac12$ boundary modes localized at the left and right internal interfaces. When the interface separation \(R\) is small, these two modes strongly overlap and form a single broad central structure. As \(R\) increases, they evolve into two well-separated localized packets centered at the two internal interfaces. The resulting low-energy manifold is therefore the direct analogue of the familiar boundary-spin manifold of a finite Haldane chain, except that the relevant spin-$\tfrac12$ modes reside at engineered internal interfaces rather than at physical open ends.

Within this manifold, the lower state \(\langle S_i^z\rangle^{(a)}\) may remain only weakly polarized throughout the chain, even though the two interface modes are already present. This occurs because the two internal spin-$\tfrac12$ degrees of freedom can combine into a low-spin configuration whose net local \(\langle S_i^z\rangle\) is strongly suppressed. By contrast, in the higher state \(\langle S_i^z\rangle^{(b)}\) the same interfacial modes become visibly polarized, producing the two localized packets seen in the main text. The near-vanishing of \(\langle S_i^z\rangle^{(a)}\) should therefore not be interpreted as absence of the interface modes, but rather as a weakly polarized representative of the same low-energy manifold.

The difference between the two sector-selected states is captured more directly by the splitting
\begin{equation}
\Delta(R)=E^{(b)}-E^{(a)},
\end{equation}
and by the real-space evolution of the polarized response in \(\langle S_i^z\rangle^{(b)}\). Together, these quantities show that the two internal modes are present already in the low-energy manifold and that their coupling decays approximately exponentially with separation.

\begin{figure}[!ht]
\centering
\includegraphics[width=\textwidth]{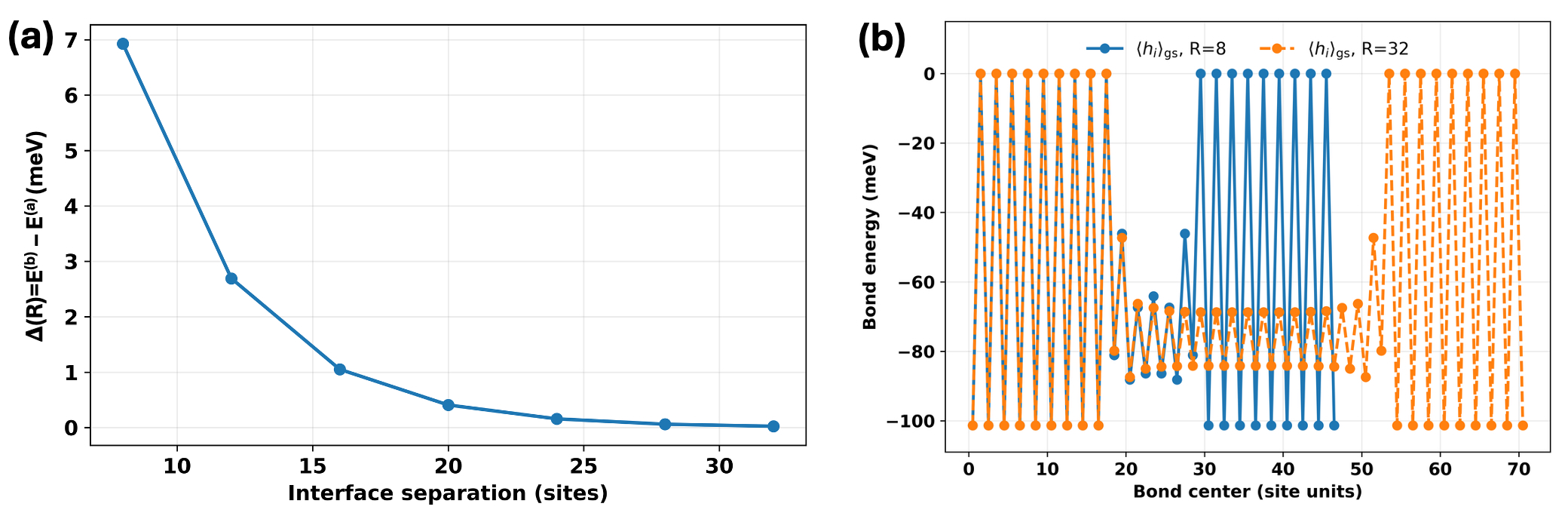}
\caption{\textbf{Additional splitting and bond-energy diagnostics for the two-interface \(A|B|A\) geometries.}
(a) Low-energy splitting \(\Delta(R)=E^{(b)}-E^{(a)}\) plotted on a linear scale as a function of interface separation \(R\). (b) Representative bond-energy profiles \(\langle h_i\rangle\) for short and long separations. The rapid suppression of \(\Delta(R)\) and the real-space separation of the bond reconstruction support the interpretation that the two internal boundary modes of the embedded effective Haldane domain hybridize strongly at short distance and decouple progressively as their separation increases.}
\label{fig:figs2}
\end{figure}

\section{S12. Additional two-interface diagnostics}

Figure \ref{fig:figs2} provides additional diagnostics for the \(A|B|A\) two-interface geometries discussed in the main text. In the main manuscript, the central result is the approximately exponential decay of the low-energy splitting
\begin{equation}
\Delta(R)=E^{(b)}-E^{(a)}
\end{equation}
with the separation \(R\) between the two internal interfaces. Fig.~\ref{fig:figs2}(a) shows the same splitting on a linear scale. The rapid suppression of \(\Delta(R)\) with increasing \(R\) is already evident before taking the logarithm used in the main text, demonstrating directly that the interaction between the two internal boundary modes becomes very weak once the embedded effective Haldane domain is sufficiently wide.

The associated local bond reconstruction is shown in Fig.~\ref{fig:figs2}(b) for representative short- and long-separation geometries. For small \(R\), the bond-energy pattern in the embedded Haldane domain remains strongly influenced by both interfaces simultaneously, indicating substantial overlap between the two internal boundary modes. For large \(R\), the local reconstruction near the left and right interfaces becomes well separated, while the middle of the Haldane domain approaches its own characteristic bond texture. This real-space evolution is fully consistent with the splitting analysis: the two internal modes strongly overlap at short separation, but become progressively decoupled as \(R\) increases.

Taken together, Figures \ref{fig:figs1} and \ref{fig:figs2} provide bond-energy and entanglement diagnostics complementary to the spin-texture analysis in the main text. They support the interpretation that the single-interface physics is termination-controlled and that the two-interface splitting originates from hybridization of localized internal boundary modes of the embedded effective Haldane segment.

\bibliographystyle{apsrev4-2}